\documentclass[12pt,epsf]{article}
\usepackage{amsmath}
\usepackage{color}
\usepackage{axodraw}
\usepackage{amssymb}
\usepackage{graphics}
\input epsf

\begin{document}
\baselineskip 24pt

\newcommand{\be}{\begin{equation}}
\newcommand{\ee}{\end{equation}}

\newcommand{\bea}{\begin{eqnarray}}
\newcommand{\eea}{\end{eqnarray}}
\newcommand{\nn}{\nonumber}
\newcommand{\eq}{\ref}
\newcommand{\ol}{\overline}

\newcommand{\sheptitle}
{Small Yukawa Couplings from Type I String Theory and the 
Inflationary Solution to the Strong CP and $\mu$ Problems}

\newcommand{\shepauthor}
{O. J. Eyton-Williams and S. F. King}

\newcommand{\shepaddress}
{School of Physics and Astronomy, University of Southampton, \\
        Southampton, SO17 1BJ, U.K.}
\vspace{0.25in}

\newcommand{\shepabstract} {We investigate the origin of
phenomenologically interesting small Yukawa couplings in Type I string
theory.  Utilising the framework of intersecting sets of D9 and
orthogonal D5 branes we demonstrate the connection between extra
dimensional volumes and Yukawa couplings. For example, we show that extra
dimensions with inverse lengths of $10^8$ GeV can lead to $10^{-10}$
Yukawa couplings.  String selection rules, arising from the D-Brane
setup, impose non-trivial constraints on the set of allowed
superpotentials.  As a phenomenological application of these 
results we construct a type I string model of inflationary particle physics
which involves small Yukawa couplings of order $10^{-10}$, and 
simultaneously solves the strong CP and $\mu$ problem
of the MSSM, via the vacuum expectation value of the inflaton field.}

\begin{titlepage}
\begin{flushright}
hep-ph/0502156 \\
\end{flushright}
\begin{center}
{\large{\bf \sheptitle}}
\\ \shepauthor \\ \mbox{} \\ {\it \shepaddress} \\
{\bf Abstract} \bigskip \end{center} \setcounter{page}{0}
\shepabstract
\begin{flushleft}
\today
\end{flushleft}

\vskip 0.1in
\noindent

\end{titlepage}

\newpage

\section{Introduction} \label{sec:Introduction}
Small Yukawa couplings are ubiquitous in the the flavour sector 
of the Standard Model describing the light fermion masses
arising from the Higgs mechanism. An extreme example of such small
Yukawa couplings is provided by Dirac neutrinos which would
require Yukawa couplings of order $10^{-12}$ in order to account for 
the very small masses. Small Yukawa couplings also arise
in other theories beyond the Standard Model, for example some
theories of inflation where the inflaton potential is required to 
be very flat. The purpose of this paper is to discuss the origin
of small Yukawa couplings in the framework of type I string theory,
and to describe an application of the results to an inflationary 
particle physics model
which involves small Yukawa couplings of order $10^{-10}$, and 
simultaneously solves the strong CP and $\mu$ problem
of the MSSM, via the vacuum expectation value of the inflaton field.

Type I string theory typically also involves the notion of
intersecting Dirichlet-branes (D-branes) 
\cite{Johnson:2000ch, Polchinski:1996na}.
The study of such frameworks has provided both a powerful
model-building tool and an elegant intuitive picture of the resulting
``brane worlds''. The particles
seen in nature are the lowest energy excitations of strings stretching
between branes.  They transform as chiral representations of the gauge
groups that correspond to the coincident stacks of D-branes.  The
stacks of D-branes considered in this paper are both D9-branes, that fill
spacetime, and three sets of D5-branes, that wrap 2-tori.  The
D5-brane stack's intersection is Minkowski space and they are orthogonal in the
extra dimensions.  When the spatial dimensions are reduced to four, by
orbifold compactification, the gauge and Yukawa couplings of each set
of branes become functions of the dimensions they wrap
\cite{Ibanez:1998rf}.  It is the connection between dimensions and
couplings that we investigate in this paper.  {\em The crucial point for
model building is the fact that small Yukawa couplings are readily
achievable in this framework}.  We discuss the conditions under which
these small couplings can arise and construct a phenomenologically
viable model utilising these couplings.  The model in question was
recently proposed in \cite{Eyton-Williams:2004bm} as a field theory model involving small 
Yukawa couplings of order $10^{-10}$
 and provides an
inflationary solution to the $\mu$ and strong CP problems. 
Our main focus here is in the string construction of the model,
and in particular on
the origin of the small Yukawa coupling present in the model.

The outline of the rest of the paper is as follows.  Section
\ref{sec:Couplings} examines the origins of the string
superpotential and the couplings in the theory.  We review the field
theory model in section \ref{sec:TheModel}. Then, in subsection
\ref{sec:StringEmbed}, we demonstrate how the model could be embedded in
string theory.  Subsection \ref{sec:Soft} introduces the string
treatment of the soft terms and provides motivation for their
spectrum.  In subsection \ref{sec:Twisted} we use new formulae for the
soft terms in the presence of twisted moduli and show that there exist
solutions consistent with the requirements of our model.  Section
\ref{sec:Conclusions} concludes the paper.

\section{Couplings and Dilaton/Moduli} \label{sec:Couplings}
We will now review the properties of Type I string theory relevant for model building, first presented in \cite{Ibanez:1998rf}. We will be working with a D-brane setup which includes a geometric mechanism for generating small gauge and Yukawa couplings.   We consider the class of spaces known as orientifolds (see \cite{Aldazabal:1998mr} for a study of possible orientifolds) requiring the addition of intersecting stacks of orthogonal D5-branes and space-filling D9-branes for consistency.  These spaces are all constructed from a 6-torus and it is the volume and anisotropy of this torus that leads to the generation of a hierarchy of couplings. The 6-torus itself is constructed out of three 2-tori each of which has one radius associated with it. We will show that if one radius is of order $10^{-8}$ $\mbox{GeV}^{-1}$ and the other two are of order $10^{-18}$ $\mbox{GeV}^{-1}$ then we obtain a coupling of order $10^{-10}$.

After compactification we end up with, in the most general case, a model consisting of three orthogonal stacks of D5-branes and a stack of D9-branes.  Each D5-brane wraps around a 2-torus and its gauge coupling depends on the radius of the torus, via the moduli, $T_i$.  The
D9-branes wrap all the tori and so depend on all the radii, via the
four dimensional dilaton, $S$.  The moduli and dilaton take the
following forms:
\begin{align}
  T_i=\frac{2R_i^2 M_*^2}{\lambda_I} +i\eta_i \label{eq:Ti}
\end{align}
\begin{align}
  S=\frac{2R_1^2 R_2^2 R_3^2 M_*^6}{\lambda_I} +i\theta \label{eq:S}
\end{align}
where $\eta_i$ and $\theta$ are untwisted Ramond-Ramond closed string states, $M_*$ is the string scale and $\lambda_I$ is the ten dimensional dilaton which governs the strength of string interactions.

The gauge couplings on the branes can be determined from the $S$ and
$T_i$ fields by 
\begin{align}
  g_{5_i}^2=\frac{4\pi}{Re(T_i)} \label{eq:g5i}
\end{align}
\begin{align}
  g_{9}^2=\frac{4\pi}{Re(S)} \label{eq:g9}
\end{align} 
The non-canonical, $D=4$, $\mathcal{N}=1$ effective superpotential has only $\mathcal{O}(1)$ Yukawa couplings \cite{Everett:2000up}, but the Kahler metric, although diagonal, is significantly different from the identity. To understand our theory in the low energy, after the dilaton and moduli have acquired VEVs, we must canonically normalise the Kahler potential and take the flat limit in which $M_p \rightarrow \infty$ while $m_{3/2}$ is kept constant \cite{Brignole:1997dp, Kaplunovsky:1993rd}.  This gives a theory containing superfields with canonical kinetic terms interacting via renormalisable operators.  Notice that the Yukawa couplings can be identified with the gauge couplings (up to the $\mathcal{O}(1)$ factors present before normalisation):
\begin{align}
 W = g_9 & \left(C_1^9 C_2^9 C_3^9 + C^{5_1 5_2} C^{5_2 5_3} C^{5_3 5_1}+
  \sum_{i=1}^3 C_i^9 C^{95_i}C^{95_i} \right)
  +\sum_{i,j,k=1}^3g_{5_i}\left(C_1^{5_i}C_2^{5_i}C_3^{5_i}
  \right. \nonumber\\
 &  +  C_i^{5_i}C^{95_i}C^{95_i} + d_{ijk}C^{5_i}_j C^{5_i 5_k}
 C^{5_i 5_k} + \frac{1}{2}d_{ijk}C^{5_j 5_k} C^{95_j} C^{95_k})
 . \label{eq:W}
\end{align}
where the $C$ terms are low energy excitations of strings: charged chiral superfields.  The superscripts denote the branes which the strings end on and terms with different subscripts transform differently under the gauge group associated with the brane.  Appendix \ref{sec:StringAssignments} discusses these fields in more detail.

Associated with this superpotential there are a set of allowed soft breaking terms given in
\begin{align}
  V_{soft} = g_9\left(A_{C^9_1 C^9_2 C^9_3}C^9_1 C^9_2 C^9_3  +  A_{C^{5_1 5_2} C^{5_2 5_3} C^{5_3 5_1}} C^{5_1 5_2} C^{5_2 5_3} C^{5_3 5_1} \right. \nonumber \\
   +  \sum_{i=1}^3\left. A_{C^9_i C^{95_i} C^{95_i}} C^9_i C^{95_i} C^{95_i} \right) + \sum_{i,j,k=1}^3 g_{5_i}\left(A_{C_1^{5_i}C_2^{5_i}C_3^{5_i}}C_1^{5_i}C_2^{5_i}C_3^{5_i} \right. \nonumber \\
   + A_{C_i^{5_i}C^{95_i}C^{95_i}} C_i^{5_i}C^{95_i}C^{95_i} + A_{C^{5_i}_jC^{5_i 5_k}C^{5_i 5_k}} d_{ijk} C^{5_i}_jC^{5_i 5_k}C^{5_i 5_k} \nonumber \\ 
   + \frac{1}{2} d_{ijk} \left. A_{C^{5_j 5_k} C^{95_j} C^{95_k}} C^{5_j 5_k} C^{95_j} C^{95_k}\right) \nonumber \\
   + \sum_{i=1}^3 m_{C^9_i}^2|C^9_i|^2 + \sum_{i=1}^3 \left(m_{C^{5_i}_1}^2|C^{5_i}_1|^2 + m_{C^{5_i}_2}^2|C^{5_i}_2|^2 + m_{C^{5_i}_3}^2|C^{5_i}_3|^2\right) \nonumber \\
   + \frac{1}{2}\sum_{i,j=1}^3 d_{ij} m_{C^{5_i 5_j}}^2|C^{5_i 5_j}|^2 +\sum_i^3 m_{C^{95_i}}^2|C^{95_i}|^2
\end{align}
where, in a slight abuse of notation, we have used the same notation for the superfields and their scalar components.  Also we have defined $d_{ijk}= |\epsilon_{ijk}|$ and $d_{ij}=|\epsilon_{ij}|$.

The $D=4$ Planck scale is related to the string scale by
\begin{align}
  M_p^2=\frac{8M_*^8 R_1^2 R_2^2 R_3^2}{\lambda_I^2}.
\end{align}
From this and Eqs.~(\ref{eq:S}) to (\ref{eq:g9}) we find that
\begin{align}
  g_{5_1}g_{5_2}g_{5_3}g_{9}=32\pi^2\left(\frac{M_*}{M_p}\right)^2. \label{eq:gaugereln}
\end{align}

In the phenomenological application discussed in the next section, we
shall require at least one coupling of $\mathcal{O}(10^{-10})$ and one
of $\mathcal{O}(1)$.  According to the above results, this constrains
the size of our radii and the value of the string scale.  For
definiteness we consider the case where $g_{5_1}\sim 10^{-10}$ and the
remaining gauge couplings are all $\mathcal{O}(1)$. From
Eq.~(\ref{eq:gaugereln}) we see this is clearly allowed if we have a
$10^{13}$ GeV string scale. Specifically our couplings are
$g_{5_2}=\sqrt{\frac{4\pi}{24}}$ (to give $\alpha_{\mbox{\tiny
GUT}}=1/24$, consistent with gauge coupling unification),
$g_{5_3}=g_9=2$ and $g_{5_1}=10^{-10}$ gives $M_*= 10^{13}$ GeV.

The hierarchy in gauge couplings corresponds to a hierarchy in the radii.  Using Eqs.~(\ref{eq:Ti}) and (\ref{eq:g5i}) for the above couplings we find that
\begin{align}
  R_1^{-1} = 1.3 \times 10^8 \mbox{ GeV} \\
  R_2^{-1} = 9.1 \times 10^{17} \mbox{ GeV} \label{eq:R2} \\
  R_3^{-1} = 2.4 \times 10^{18} \mbox{ GeV}.
\end{align}
These radii are all too small to have Kaluza-Klein (KK) or winding modes that will be readily excitable at collider energies. The winding modes of $R_1$ are $\approx n 10^{18}$ GeV and $R_2$ and $R_3$ have winding modes of $\approx n 10^{8}$ GeV.  The KK modes for $R_1$ are $\approx n 10^8$ GeV and $R_2$ and $R_3$ are $\approx n 10^{18}$ GeV. In principle these massive modes could affect inflation.  However the inflationary scale is $10^8$ GeV so it is unlikely that these modes would appear with any great abundance.

There is a caveat to Eq.~(\ref{eq:gaugereln}). It is only valid when
the untwisted moduli provide the dominant contribution to the gauge
couplings.  This will be true in the application discussed in the
next section, to a reasonable approximation.
For large (negative) values for $\delta_{GS}$ (see \cite{King:2001zi})
we need large twisted moduli (to be discussed in Section \ref{sec:Twisted}) on the
Standard Model brane, D$5_2$, so the twisted modulus has a significant
contribution to $g_{5_2}$.  In this case it is better to consider the
alternative formulation of Eq.~(\ref{eq:gaugereln}) in terms of
$S/T_i$:
\begin{align}
  \left(\frac{M_*}{M_p}\right)^2=\frac{1}{2 \bigl(Re(T_1)Re(T_2)Re(T_3)Re(S)\bigr)^{1/2}}
\end{align}
From this we can see that $M_*$ only has a very weak dependence on $Re(T_2)$, $M_*\propto \bigl(Re(T_2)\bigr)^{-1/4}$.  As such doubling $Re(T_2)$ only amounts to a 20\% correction to $M_*$.  Since the exact value of $M_*$ is not crucial to our results we use the approximate formula Eq.~(\ref{eq:gaugereln}) exclusively in this paper.

It is worth noting that we could have chosen any of the three sets of five branes to have the tiny coupling, but D9 would not have been a good choice because of the following relation\footnote{Again there is an alternative formulation in terms of the moduli and dilaton: $\lambda_I^2 = 2 Re(S)/\bigl(Re(T_1)Re(T_2)Re(T_3)\bigr)$}.
\begin{align}
  \lambda_I = \frac{g_{5_1} g_{5_2} g_{5_3}}{2\pi g_9}
\end{align}
To remain in the perturbative regime we require that $\lambda_I$ be less than one.  Clearly, swapping between ($g_9\sim 10^{-10}$, $g_{5_1}\sim 1$) and ($g_9\sim 1$, $g_{5_1}\sim 10^{-10}$) dramatically changes $\lambda_I$, assuming all the other couplings are left unchanged. The gauge coupling choices shown below Eq.~(\ref{eq:gaugereln}) give $\lambda_I\sim 10^{-11}$ which is well inside the perturbative regime.  It should be noted that, under T-Duality, these two cases are equivalent.  However, if we want to perform string theoretic calculations a perturbative coupling is undeniably useful.

\section{A Phenomenological Application} \label{sec:TheModel}
\subsection{The Inflationary Solution to the Strong CP and $\mu$ Problems}
As a phenomenological application of these results we shall 
discuss the supersymmetric field theory model proposed in
\cite{Eyton-Williams:2004bm} where small Yukawa couplings 
$\lambda, \kappa$ of order
$10^{-10}$ were invoked. To make this paper self-contained we briefly review
the field theory model, are we refer the reader to
\cite{Eyton-Williams:2004bm} for a more detailed discussion. 
The idea of the model was to provide a simultaneous solution to both
the $\mu$ problem and the strong CP problem, as well as providing 
a satisfactory model of hybrid inflation. Indeed the role of the
inflaton $\phi$ is rather special in this model, since it not only
provide inflation, but also its vacuum expectation value after inflation
is directly responsible for the $\mu$ term of the MSSM.

The starting point of the field theory model is the following superpotential and potential:
\begin{align}
  W=\lambda \phi H_u H_d + \kappa \phi N^2 + y_t Q H_u U + y_b Q H_d D. \label{eq:NewW}
\end{align}
\begin{align}
  V_{soft} = & \lambda A_\lambda \phi H_u H_d + \kappa A_\kappa \phi N^2 +  y_t A_{y_t} Q H_u U + y_b A_{y_b} Q H_d D + h.c. \nonumber \\ 
  &  + m_0^2(|H_u|^2 + |H_d|^2 + |N|^2 + |U|^2 + |D|^2) + m_Q^2|Q|^2 + m_\phi^2|\phi|^2 \label{eq:NewV}
\end{align}
$\phi$ and $N$ are, respectively, the inflaton
and waterfall fields (for a more detailed discussion of the
inflationary epoch see \cite{Eyton-Williams:2004bm} and the references
therein).  These fields are singlets of the Minimal Supersymmetric
Standard Model (MSSM) \cite{Chung:2003fi} gauge group and the other
fields are just the usual quarks and Higgs multiplets of the MSSM
with standard MSSM quantum numbers as shown in table \ref{tab:MSSM}.

\begin{table}[h!]
  \begin{equation*}
    \begin{array}{|c|c|c|c|}
      \hline
      \mbox{Fields}  & SU(3) & SU(2)_L & U(1)_Y \\
      \hline
      \hline
      Q & 3 & 2 & 1/6 \\
      \hline 
      U & \bar{3} & 1 & -2/3 \\
      \hline
      D & \bar{3} & 1 & 1/3 \\
      \hline
      H_u & 1 & 2 & 1/2 \\
      \hline
      H_d & 1 & 2 & -1/2 \\
      \hline
    \end{array}
  \end{equation*}
  \caption{MSSM charges} \label{tab:MSSM}
\end{table}

 This is a model of hybrid inflation
 \cite{Linde:1990gz,Linde:1991km,Copeland:1994vg,Lyth:1996we,Linde:1997sj}
 that simultaneously solves the $\mu$ \cite{Chung:2003fi} and strong
 CP problems.  Hybrid inflation is a two field inflationary model with
 one field, the inflaton, that drives inflation and another field, the
 waterfall field, that ends inflation.  The coupling of these two
 fields provides a field-dependent mass for the waterfall field which
 becomes tachyonic at a critical value of the inflaton, triggering the
 end of inflation.  After inflation has ended the VEV of the inflaton
 $<\phi>$ both generates the $\mu$ term (in a similar way to the NMSSM
 \cite{NMSSM, NMSSMphenom}), when $\lambda \phi H_u H_d \rightarrow
 \mu H_u H_d$ and breaks Peccei-Quinn (PQ) \cite{Peccei:1977hh} symmetry
 solving the strong CP problem. The PQ breaking scale was shown
to be $<\phi>\sim 10^{13} \ {\rm GeV}$, and $\lambda \sim 10^{-10}$
then results in $\mu \sim 10^3 \ {\rm GeV}$. Inflationary
 considerations also require $\kappa \sim 10^{-10}$.

This model is similar to the BGK model \cite{Bastero-Gil:1997vn}, but
with the inflaton providing the $\mu$ term rather than the $N$ field.
In \cite{Eyton-Williams:2004bm} it was assumed {\it ad hoc} that
$\lambda = \kappa$ and $A_{\lambda}= A_{\kappa}$, where both of these
results will shortly be derived from the string construction.
This led to the prediction \cite{Eyton-Williams:2004bm}:
\begin{align}
  8m_0^2>|A_\lambda|^2>4m_0^2 \label{eq:bounds}
\end{align}
and hence:
\begin{align}
  \mu^2=(0.25-0.5)m_0^2,
\end{align}
where $m_0$ is a universal soft scalar mass of order a TeV,
whose universality will also be shown to result from the string
construction. The relative smallness of the inflaton soft scalar 
mass, which is assumed to be of order an MeV, will also be discussed.

\subsection{String Embedding of the Model} \label{sec:StringEmbed}
  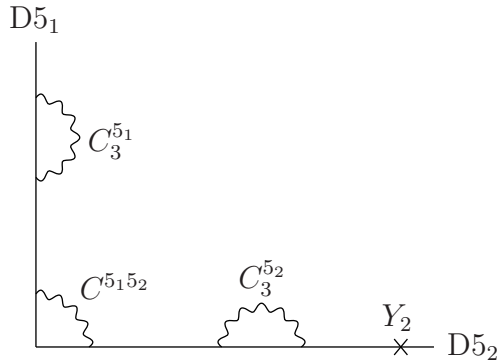
\begin{figure}[h!]
  \begin{picture}(300,125)(0,0)
    \Line(100,10)(250,10)
    \Text(256,10)[l]{D$5_2$}
    \Line(100,10)(100,125)
    \Text(100,134)[c]{D$5_1$}
    \PhotonArc(100,10)(20,0,90){1.5}{4.5}
    \PhotonArc(185,10)(15,0,180){1.5}{6.5}
    \PhotonArc(100,89)(15,-90,90){1.5}{6.5}
    \Line(235,13)(240,7)
    \Line(240,13)(235,7)
    \Text(237,21)[c]{$Y_2$}
    \Text(117,33)[l]{$C^{5_1 5_2}$}
    \Text(186,36)[c]{$C^{5_2}_3$}
    \Text(120,88)[l]{$C^{5_1}_3$}
  \end{picture} \caption{\small Schematic representation of two stacks of D5-branes.  The stacks of branes overlap in Minkowski space, but are orthogonal in the compactified dimensions. The $C$ states correspond to chiral matter fields and $Y_2$ is a twisted modulus (introduced in Section \ref{sec:Twisted}) localised within the extra dimensions, but free to move in Minkowski space.  We have only presented the string states involved in our construction: for a more complete picture see Figure 1 in \cite{King:2001zi}.} \label{fig:one}
  \end{figure}
  Now that we have the superpotential Eq.~(\ref{eq:W}) and the Yukawa
  relationship, Eq.~(\ref{eq:gaugereln}), we have all the tools
  necessary for the string construction.  Figure~\ref{fig:one}
  displays the two branes that feature in our construction. The D$5_2$
  brane is assumed to have an order one gauge coupling and possesses a
  twisted moduli on the fixed point that is spatially separated from
  the D$5_1$ brane.  All the MSSM fields transform under
  representations of the D$5_2$ brane's gauge group.  The D$5_1$ brane
  is assumed to have an order $10^{-10}$ gauge coupling and hence no
  MSSM fields transform under it.  
This corresponds to the choice of gauge couplings already discussed
in section~\ref{sec:Couplings}.

The specific string assignments we
assume are given in table~\ref{tab:Assignments}.
\begin{table}[h]
    \begin{equation*}
      \begin{array}{|c|c|c|c|c|c|c|}
        \hline
        \phi & N & Q & H_u & H_d & U & D \\ 
        \hline
        \hline
        C^{5_1}_3 &  C^{5_1 5_2} & C^{5_2}_3  & C^{5_1 5_2} & C^{5_1 5_2} & C^{5_1 5_2} & C^{5_1 5_2}  \\
        \hline
      \end{array}
    \end{equation*}  \caption{\small String Assignments} \label{tab:Assignments}
  \end{table}
\begin{table}[h]
  \begin{equation*}
    \begin{array}{|c|ccc||c|ccc|}
      \hline
      g_{5_1} & \multicolumn{3}{|c|} {C^{5_1}_3 C^{5_1 5_2} C^{5_1 5_2}} & g_{5_2} & \multicolumn{3}{|c|}{C^{5_2}_3 C^{5_2 5_1} C^{5_2 5_1}} \\
      \hline
      \hline
      \lambda & \phi & N & N & y_t & Q & H_u & U \\
      \kappa & \phi & H_u & H_d & y_b & Q & H_d & D \\
      \hline
    \end{array}
  \end{equation*}  \caption{\small String allowed Yukawa couplings arising
      from the string assignments in table \ref{tab:Assignments}.} \label{tab:Alternative Assignments}
\end{table}

The allowed couplings which result from the assumed string assignments
follow from the string selection rules embodied in the superpotential
in Eq.~(\ref{eq:W}), and 
are written explicitly in table~\ref{tab:Alternative Assignments}.
Table \ref{tab:Alternative Assignments} demonstrates that, with the
assumed string assignments, we can reproduce the field theory model
superpotential in Eq.~(\ref{eq:NewW}). 
We emphasise that we now have a stringy
explanation for the small Yukawa couplings in the field theory model,
since in the string construction 
they are equal to the small gauge couplings $g_{5_1}\sim 10^{-10}$ whose
smallness has a string origin as discussed in
section~\ref{sec:Couplings}.
Note that the allowed couplings in table
\ref{tab:Alternative Assignments} are precisely the ones assumed in
the superpotential, Eq.~(\ref{eq:W}), and gauge invariance means that
the only terms we are allowed to write down are those found in
Eq.~(\ref{eq:NewW}), and no others.
Furthermore the string selection rules now require
that $\lambda = \kappa = g_{5_1}\sim 10^{-10}$, 
and $y_t=y_b=g_{5_2} \sim 1$, previously assumed in an {\it ad hoc}
manner in the field theory version of the model.

Table~\ref{tab:Alternative Assignments} also
makes clear that $g_{5_1} C^{5_1}_3 C^{5_1 5_2} C^{5_1
5_2}$ and $g_{5_2} C^{5_2}_3 C^{5_2 5_1} C^{5_2 5_1}$ cannot produce
any further gauge invariant superpotential terms.  Also, since we have
only assigned fields to $C^{5_1}_3$, $C^{5_2}_3$ and $C^{5_1 5_2}$ the
only terms in Eq.~(\ref{eq:W}) that we can use are the two
superpotential terms that appear in table \ref{tab:Alternative
Assignments}.  It should be highlighted that the assignment of quark
Yukawas is non-trivial if there is a direct connection (i.e. a
renormalisable coupling) between the inflationary interactions and
MSSM fields.  For example, if the $\mu$ term were provided by the $N$
field, as in \cite{Bastero-Gil:1997vn}, it would not be possible to
write down quark Yukawas consistent with the inflationary sector of
the model.

The methodology underlying the selection of these assignments is the subject of a detailed discussion in Appendix \ref{sec:Methodology}.

\subsection{Soft Terms} \label{sec:Soft}
Given the assumption that the SUSY breaking is dominated by the
dilaton/moduli sector we can parametrise its effect in terms of
$m^2_{3/2}$ and Goldstino angles (that parametrise the contributions from different sources of SUSY breaking). The following, Goldstino angle
independent, sum rule is crucial to our model as we will soon see:
\begin{align}
  m^2_{C^{5_1}_3} + 2 m^2_{C^{5_1 5_2}}=|M_{5_1}|^2=|A_{C^{5_1}_3 C^{5_1 5_2} C^{5_1 5_2}}|^2 \label{eq:tri}
\end{align}

In order to obtain successful slow roll \cite{Lyth:1998xn} we need to
have a small soft mass for the inflaton, $\phi$, (MeV or less), which
presents us with a potential difficulty.  All the soft mass squareds
coming from the dilaton/moduli SUSY breaking are, assuming vanishing
cosmological constant, of the form $m^2_{3/2} F(\theta,\Theta_i)$
where $F$ is some, typically order one, function of the Goldstino
angles, $\theta, \Theta_i$.  If we appeal to Gaugino condensation we
can motivate the selection of Goldstino angles resulting in a small
mass for $\phi$ since this allows explicit calculation of the soft
terms. In \cite{Abel:2000tf} the soft spectrum resulting from gaugino
condensation was shown to include $m^2_{C^{5_1}_3}\approx 0$ and
$m^2_{C^{5_1 5_2}} \approx \frac{1}{2}m^2_{3/2}$.  From (\ref{eq:tri})
we see that $|A_{C^{5_1}_3 C^{5_1 5_2} C^{5_1 5_2}}|^2 \approx
m^2_{3/2}$.  The validity of the sum rule is noted in their paper and
they present a consistent, explicit expression for the trilinear
couplings.

It is worth mentioning that this is true for orbifold compactifications in the absence of twisted moduli, but not for more general spaces, such as Calabi-Yau manifolds \cite{Kim:1996wh}, but for the purposes of this paper we restrict ourselves to orbifolds.

Now we must consider the implications of this sum rule for our model.  From Eq.~(\ref{eq:tri}) we have an explicit relationship between the soft masses and trilinears for the relevant string assignments. From table \ref{tab:Assignments} we know that $\phi$ is assigned to $C^{5_1}_3$ and that $N$ is assigned to $C^{5_1 5_2}$.  Since $m_\phi\approx 0$ and $N$, $H_u$ and $H_d$ have the same string assignment Eq.~(\ref{eq:tri}) becomes
\begin{align}
  2m^2_0\approx |A_\lambda|^2,
\label{approxsumrule}
\end{align}
where the string prediction is that $A_\lambda = A_\kappa = 
A_{C^{5_1}_3 C^{5_1 5_2} C^{5_1 5_2}}$, and 
$m_0^2=m^2_{C^{5_1 5_2}}$, where we have assumed that
$m_\phi^2 =m^2_{C^{5_1}_3}\approx 0$.
Note that in \cite{Eyton-Williams:2004bm} it was assumed {\it ad hoc} that
$A_{\lambda}= A_{\kappa}$, but now this result follows immediately
from the string construction.

Unfortunately Eq.~(\ref{approxsumrule}) 
does not satisfy the lower bound on the trilinears shown in
Eq.~(\ref{eq:bounds}) so the standard soft terms are inconsistent with
our model.  Since the soft terms depend, via the SUGRA potential, on
the F-terms and the Kahler potential \cite{Kaplunovsky:1993rd} we must
look to modifying these. In fact in type I string theory it is known
that there are in general twisted moduli which, if their F-terms
have non-zero VEVs will lead to additional contributions to 
supersymmetry breaking described by extra Goldstino angles 
\cite{King:2001zi}, which will lead to a 
violation of the above sum rule for soft masses.
In the following section we consider the
Kahler potential in the presence of one twisted modulus located at a
fixed point in the 2-torus wrapped by D$5_2$.

\subsection{Soft terms with Twisted Moduli} \label{sec:Twisted}
We want to quantify the effects of twisted moduli on our string set up
and see if it is possible to include them in such a way as to give
acceptable phenomenology. We follow the analysis in
\cite{King:2001zi}, but generalise it to allow $\lambda_I\neq 1$. Our
work will show that this produces soft terms consistent with those
presented in \cite{King:2001zi} when $\lambda_I=1$, but that
$\lambda_I$ has a significant effect on the soft spectrum.  We
consider the case of just one twisted modulus, $Y_2$, living on D$5_2$
at the fixed point shown on figure \ref{fig:one}.  Leaving the
derivation of the modified soft terms to Appendix \ref{sec:SoftApp},
we quote the important results here.  First the soft masses:
\begin{align}
  m_Q^2 & = m_{C^{5_2}_3}^2 = m_{3/2}^2 - 3m_{3/2}^2 \Theta_1^2 \cos^2\theta \sin^2\phi \nonumber\\
  m_0^2 & = m_{C^{5_{1} 5_{2}}}^{2} = \tilde{m}^{2} - \frac{3}{2} m_{3/2}^{2} 
  \left( \sin^{2} \theta + \Theta_{3}^{2} \, \cos^{2} \theta \, \sin^{2} \phi 
  \right)  \nonumber \\
  m_\phi^2 & = m_{C^{5_{1}}_{3}}^{2} = \tilde{m}^{2} - \frac{3}{k} m_{3/2}^{2}  \Theta_{2}^{2} \cos^{2}\theta \sin^{2}\phi. \label{eq:masses}
\end{align}
where Goldstino angles $\theta$ and $\phi$ and Goldstino parameters
$\Theta_i$ (where $\sum_{i=1}^3 \Theta_i^2=1$) parametrise the
supersymmetry breaking as discussed in \cite{King:2001zi}. $\tilde{m}$ contains all the $\lambda_I$ dependence and
\begin{align}
  k=1+\delta_{GS}\left(Y_2 + \overline{Y}_2 - \delta_{GS} \ln (T_2 + \overline{T}_2)\right). \label{eq:K}
\end{align}

In full generality $\tilde{m}$ is an unwieldy expression, to be found
in Appendix \ref{sec:SoftApp}, but when $\lambda_I \ll 1$, it
simplifies considerably.  To very good approximation
$\tilde{m}\approx m_{3/2}$ in our model.  The full expression for the
trilinear associated with the inflationary superpotential terms is
quoted in Appendix \ref{sec:SoftApp}.  The approximate result when
$\lambda_I=10^{-11}$ is
\begin{align}
 A_\lambda = A_{C^{5_1}_3 C^{5_1 5_2} C^{5_1 5_2}} 
    \approx  -\sqrt{3} m_{3/2} \, \cos\theta \left[ \sin\phi \, \Theta_1 \, 
     e^{i\alpha_1} \right. \nonumber \hspace*{1.8cm} \\
 \left. - \cos\phi \, e^{i\alpha_{Y_2}} 
   \left\{ Y_2 + \overline{Y}_2 - \delta_{GS} \ln(T_2 + \overline{T}_2) \right\} 
    \right] \hspace*{1cm} \label{eq:firstaii13}
\end{align}
We now need to analyse these expressions to see if it is possible to
satisfy the bounds laid out in section \ref{sec:TheModel}.  In this
paper we do not explicitly calculate the soft terms, but rely on
\cite{Abel:2000tf} to motivate the plausibility of $m_\phi^2 \sim 0$.
For our purposes we simply wish to see if it is possible to satisfy
the bounds for a particular choice of parameters.  To see this we
have performed a Monte Carlo analysis
to generate a random set of Goldstino parameters in order to
check that the masses and trilinears fit within the allowed ranges.
In addition to the inflationary requirements we only accepted
parameters that gave positive soft mass squareds for the quark
doublet, $Q$.

Since $m_\phi^2=0$ can only be satisfied if $k$ is within the following range
\begin{align}
  0 < k \le 3 \label{eq:k range}
\end{align}
the allowed values for $Y_2 + \overline{Y}_2$ and $T_2 +
\overline{T}_2$ are restricted by Eq.~(\ref{eq:K}). 
We also wish to retain gauge coupling unification
which places further constraints on our moduli VEVs since we require
that:
\begin{align}
  \frac{4\pi}{g_\alpha^2} = Re(f_\alpha) = Re(T_2) +
  \frac{s_{\alpha}}{4\pi}Re(Y_2)\approx 24\label{eq:gaugekinetic}
\end{align}
where $s_\alpha$ are model dependent coefficients and $\alpha$ runs
over the the different gauge groups.  In some orientifolds $s_\alpha$
can be of the same order as the beta functions, as discussed in
\cite{Ibanez:1998rf}, but we make the simplifying assumption that they
are all set to $4\pi$.  The other simplifying assumption we make is to
set all the phases to 1.

The Monte Carlo routine starts with a particular value for
$\delta_{GS}$, generates a random set of Goldstino parameters, $k$ and
the moduli that satisfy gauge coupling unification and then calculates
the required masses and trilinears.  When this is complete it checks
to see if the constraints on the masses are satisfied and if they are
then the parameters are accepted.

Sample points are displayed in table \ref{tab:Data} 
with soft masses in units of $m_{3/2}$.  These sample points all
satisfy the condition in Eq.~(\ref{eq:bounds}), as required.  
Also from \cite{Eyton-Williams:2004bm} we know that, in order to satisfy slow
roll conditions, $m_\phi \sim$ MeV is required.  
In our approach this will require tuning of the Goldstino angles,
to yield the MeV inflaton soft masses, but as already mentioned
this may arise dynamically in certain approaches \cite{Abel:2000tf}. 
Note that the Monte Carlo has accurately determined the
Goldstino angles necessary to achieve such MeV soft masses explicitly,
but in table \ref{tab:Data} they have been rounded for brevity.

\begin{table}[h!]
$\begin{array}{|c|c|c|c|c|c|c|c|c|c|c|c|}
  \hline
  \delta_{GS} & \theta & \phi & \Theta_1 & \Theta_2 & \Theta_3  & m_0^2 & m_Q^2 & A_\lambda & A_Q & Re(T_2) & Re(Y_2) \\
  \hline
  -2 & 5.69 & 5.33 & 0.807 & 0.173 & 0.565 & 0.312 & 0.113 & 1.34 & 1.40 & 27.9 & -3.92 \\
  \hline
  -2 & 5.54 & 5.25 & 0.810 & 0.497 & 0.311 & 0.248 & 0.215 & 1.11 & 1.23 & 27.8 & -3.84 \\
  \hline
  -4 & 6.14 & 5.20 & 0.463 & 0.176 & 0.869 & 0.100 & 0.507 & 0.887 & 1.19 & 32.2 & -8.22 \\
  \hline
  -4 & 5.57 & 5.26 & 0.777 & 0.407 & 0.479 &  0.219 & 0.237 & 1.01 & 1.13 & 32.2 & -8.23 \\
  \hline
  -6 & 5.54 & 4.74 & 0.705 & 0.529 & 0.472 & 0.13 & 0.195 & 0.901 & 1.00 & 36.9 & -12.9 \\
  \hline
  -6 & 5.39 & 3.88 & 0.937 & 0.163 & 0.31 & 0.0679 & 0.532 & 0.551 & 0.867 & 36.8 & -12.8 \\ 
  \hline
  -8 & 6.68 & 5.19 & 0.693 & 0.156 & 0.704 & 0.279 & 0.0313 & 1.07 & 1.09 & 41.6 & -17.6 \\
  \hline
  -8 & 7.18 & 4.57 & 0.696 & 0.707 & 0.127 & 0.0724 & 0.448 & 0.735 & 0.992 & 41.7 & -17.7 \\
  \hline
  -10 & 6.4 & 4.94 & 0.379 & 0.451 & 0.808 & 0.0625 & 0.596 & 0.652 & 1.02 & 46.7 & -22.7 \\
  \hline
  -10 & 5.43 & 4.34 & 0.916 & 0.316 & 0.248 & 0.117 & 0.0505 & 0.938 & 0.963 & 46.6 & -22.6 \\
  \hline
\end{array}$ \caption{Goldstino parameters and soft terms satisfying Eq.~(\ref{eq:bounds})} \label{tab:Data}
\end{table}

\section{Conclusions}\label{sec:Conclusions}
We have discussed the possibility of generating small
Yukawa couplings within the framework of Type I string theory,
and emphasised the importance of this result for model building.
We have seen that it is possible to obtain very small
Yukawa couplings without requiring an exceptionally low string scale or
especially large extra-dimensions. In the example discussed
we obtained Yukawa couplings of order $10^{-10}$ with a string scale
of order $10^{13}\ {\rm GeV}$, and the largest extra dimensions having  
a compactification scale of order $10^{8}\ {\rm GeV}$.

As an application of these results we constructed a
type I string theory realisation of the model presented in
\cite{Eyton-Williams:2004bm} which simultaneously solves the
$\mu$ and strong CP problems while providing a viable description of
hybrid inflation. The model provides a good example of the approach
since it requires exceptionally small Yukawa couplings of order $10^{-10}$,
together with certain constraints on the soft mass parameters,
which from the field theory point of view looks quite unappealing
and {\it ad hoc}. We have shown how the model can originate from
a type I string theory setup 
consisting of intersecting D9 and D5-branes where 
the required small Yukawa couplings and soft mass relations 
can readily emerge. 
Twisted moduli play an important part in determining the 
soft spectrum, and we have provided a first discussion of their effects
away from the $\lambda_I=1$ limit.

The model discussed provides a convenient example in which 
small Yukawa couplings and soft mass relations can arise
quite naturally from a type I string construction.
It is clear that the type I string approach discussed here is much more general
than this specific model, and has very wide applicability.
A possible further application is to Dirac neutrinos, 
which also involve very small Yukawa couplings,
and we plan to discuss this in a dedicated future publication.

\appendix
\section{Methodology}\label{sec:Methodology}
There are three main rules that must be adhered to when attempting to realise field theoretic models in the D-Brane framework.  One: each superfield may only be assigned to one string state.  Two: each string state can have several different superfields assigned to it.  Three: all tree level superpotential terms have to be found in Eq.~(\ref{eq:W}).  Notice that the superpotential terms in Eq.~(\ref{eq:W}) have one of two forms: either one field linear and the other quadratic (e.g. $C^{5_i}_k C^{5_i 5_j} C^{5_i 5_j}$), or all three fields linear (e.g. $C^{5_i}_1 C^{5_i}_2 C^{5_i}_3$).  There are no cubic terms in the string superpotential.

The aim of this approach is to take a purely field theoretic superpotential and see if it can be realised in the string superpotential, using the rules as laid out.  To see the terms allowed we want to consider all possibilities in turn. First let us categorise all possible renormalisable superpotential terms that we might want to realise in the string superpotential.

For simplicity consider a toy field theory with just three, gauge singlet, superfields $A$, $B$ and $C$ and the following superpotential.
\begin{align}
  W=aABC + bAAB + cAAA \label{eq:Toy}
\end{align}
Let us demonstrate the assignment of each term in Eq.~(\ref{eq:Toy}) individually.

$aABC$ could be assigned to any term in Eq.~(\ref{eq:W}), obviously a string term of the trilinear form $C^{5_i}_1 C^{5_i}_2 C^{5_i}_3$ would be acceptable, leading to different assignments for each superfield. For example $A$, $B$ and $C$ could be assigned to $C^{5_i}_1$, $C^{5_i}_2$ and $C^{5_i}_3$ respectively.  A quadratic term like $C^9_i C^{95_i} C^{95_i}$ could also be used since $B$ and $C$ can share the same assignment, $C^{95_i}$. For example $A$ assigned to $C^{9}_i$ and both $B$ and $C$ assigned to $C^{95_i}$.  Now $AAB$ can only be assigned to quadratic terms: rule one forbids trilinear terms since we would be forced to assign one field to two string states.  We can clearly see that $AAA$ cannot be assigned to either quadratic or trilinear for the same reasons.  Incidentally this forbids the NMSSM at tree level since it requires a cubic superfield.

We have now laid out the string selection rules underlying our construction.  Obviously each term allowed by the string selection must be gauge invariant and have the correct coupling for it to appear in the theory.  Once we have made all the necessary assignments we write down all terms allowed by gauge invariance and string selection. For example, even if $c$ were set to zero, we could not get Eq.~(\ref{eq:Toy}) exactly as we would be forced to include the term $bCCB$.  With the superpotential written down and the Kahler potential canonically normalised the supersymmetric side of the construction is complete.  The SUSY breaking terms are considered earlier in the text in sections \ref{sec:Soft} and \ref{sec:Twisted}.  In the following section we explicitly construct our theory.

\subsection{String Assignments} \label{sec:StringAssignments}

First we look for all possible superpotential terms containing a squared superfield and with the correct coupling, $g_{5_1}$, to accommodate $\kappa \phi N^2$.  There are three terms that satisfy this requirement:
\begin{align}
  (i) \ g_{5_1}C^{5_1}_1 C^{9 5_1} C^{9 5_1}, (j) \ g_{5_1} C^{5_1}_2 C^{5_1 5_3} C^{5_1 5_3} \mbox{\ \ and } (k) \ g_{5_1} C^{5_1}_3 C^{5_1 5_2} C^{5_1 5_2} \label{eq:N2}
\end{align}
$(j)$ and $(k)$ are symmetric under relabelling of 2 and 3.  Notice that Eq.~(\ref{eq:W}) is symmetric under permutations of the 1, 2 and 3 labels since you are free to choose your radii at the start of the construction.  T-Duality and relabelling links all possible permutations of the branes hence $(i)$, $(j)$ and $(k)$ are equivalent at this stage of the construction.   Due to these symmetries we only consider the $C^{5_1}_3 C^{5_1 5_2} C^{5_1 5_2}$ term. Since we know that $\phi$ is assigned to $C^{5_1}_3$ we look for the $\lambda \phi H_u H_d$ term.

There are only two terms that include $\phi$: (a) $g_{5_1} C^{5_1}_3 C^{5_1 5_2} C^{5_1 5_2}$ and (b) $g_{5_1} C^{5_1}_1 C^{5_1}_2 C^{5_1}_3$.  Notice these are  inequivalent under T-Duality.

Now the possible gauge assignments become important.  Superfields transform under gauge groups according the branes they (or rather the string which they are an excitation of) stretch between.  For example a $C^{5_1}_3$ state corresponds to a string that starts and ends on the same set of branes, D$5_1$, and so can only have gauge quantum numbers from that stack.  On the other hand a $C^{5_1 5_2}$ state ends on both the D$5_1$ branes and the D$5_2$ branes and so can have quantum numbers from both.

Naturally we must ensure that our fields transform appropriately under our symmetry group, in our case the MSSM's, and that each term is invariant.  For term (b) each Higgs superfield must transform under the D$5_1$ brane's gauge group with gauge coupling $g_{5_1} \sim 10^{-10}$.  Since we expect the standard model gauge couplings to be order one at the string scale this is unacceptable.  This leaves (a): both Higgs can transform correctly since we only need one non-Abelian gauge factor, $SU(2)_L$, and this can come from the D$5_2$ brane with its order one coupling.  The final step is to see if $\mathcal{O}(1)$ quark Yukawas are consistent with these assignments.

With $H_u$ and $H_d$ both assigned to $C^{5_1 5_2}$, the order one terms we can write down are
\begin{align}
  (\alpha) g_{5_2} C^{5_2}_1 C^{5_2 5_1} C^{5_2 5_1}, \hspace{5pt} (\beta) g_9 C^{5_2 5_3} C^{5_1 5_2} C^{5_3 5_1} \mbox{ and } (\gamma) g_{5_3} C^{9 5_2} C^{5_1 5_2} C^{9 5_1}
\end{align}

To enforce gauge invariance the quark doublet must live on an assignment with ends on branes with $\mathcal{O}(1)$ gauge couplings. With this in mind the simplest choice is $(\alpha)$ since it has all the standard model gauge factors coming from one brane.  The second and third choices necessitate diagonal symmetry breaking from $\bigl(SU(3)\times SU(2) \times U(1)\bigr)^2$ to $SU(3)\times SU(2) \times U(1)$.  With this caveat all three solutions are valid models, but chose to focus on the simpler model without diagonal symmetry breaking in the rest of the analysis. This concludes the supersymmetric construction.

To compare these assignments with a T-Dual model we list the possible assignments and a T-Dual set in Appendix \ref{sec:Assignments}. 
\subsection{Alternative assignments}\label{sec:Assignments}
The following table contains all allowed assignments once $\lambda \phi N^2$ is assigned to either $C^{5_1}_1 C^{95_1} C^{95_1}$ or $C^{5_1}_3 C^{5_1 5_2} C^{5_1 5_2}$.  In the first block the couplings are $g_{9}=\sqrt{\frac{4\pi}{24}}$, $g_{5_3}=g_{5_2}=2$ and $g_{5_1}=10^{-10}$ and the second block has $g_{9}$ and $g_{5_2}$ exchanged. As pointed out in Appendix \ref{sec:StringAssignments} the two sets of assignments are equivalent.  We only include them both to allow the reader to compare and contrast both cases.
\begin{table}[h!]
  \begin{equation*}
    \begin{array}{|c|c|c|c|c|c|c|}
      \hline
      \phi & N & Q & H_u & H_d & U & D \nonumber \\
      \hline
      \hline
      C^{5_1}_1 & C^{9 5_1} & C^9_1 & C^{9 5_1} & C^{9 5_1} & C^{9 5_1} & C^{9 5_1} \\
      C^{5_1}_1 & C^{9 5_1} & C^{9 5_2} & C^{9 5_1} & C^{9 5_1} & C^{5_1 5_2} & C^{5_1 5_2} \\
      C^{5_1}_1 & C^{9 5_1} & C^{9 5_3} & C^{9 5_1} & C^{9 5_1} & C^{5_1 5_3} & C^{5_1 5_3} \\
      \hline
      C^{5_1}_3 & C^{5_1 5_2} & C^{5_2}_3 & C^{5_1 5_2} & C^{5_1 5_2} & C^{5_1 5_2} & C^{5_1 5_2}  \\
      C^{5_1}_3 & C^{5_1 5_2} & C^{5_2 5_3} & C^{5_1 5_2} & C^{5_1 5_2} & C^{5_3 5_1} & C^{5_3 5_1}  \\
      C^{5_1}_3 & C^{5_1 5_2} & C^{9 5_2} & C^{5_1 5_2} & C^{5_1 5_2} & C^{9 5_1} & C^{9 5_1}  \\
      \hline
    \end{array}
  \end{equation*} \caption{Collated assignments}
\end{table}

\section{Twisted soft terms}\label{sec:SoftApp}

To include the effects of twisted moduli the Kahler potential must be modified. The twisted moduli need a quadratic term in the Kahler potential to provide their kinetic terms:
\begin{equation}
  \hat{K}( Y_{2}, \overline{Y}_{2} ) = \frac{1}{2} \left[ Y_{2}+\overline{Y}_{2} - \delta_{GS} 
    \ln (T_{2} + \overline{T}_{2}) \right]^{2}
  \label{eq:ykahler2}.
\end{equation}
The $\delta_{GS} \ln (T_{2} + \overline{T}_{2})$ term describes anomaly cancellation via Green-Schwarz mixing \cite{Ibanez:1998qp}.

To enforce the sequestration of states the Kahler potential must be modified accordingly. A multiplicative factor $\xi$ must be introduced to every string state that does not overlap significantly with the twisted modulus.  For example a $C^{5_2}_2$ state comes from a string which starts and ends on D$5_2$ so it interacts with $Y_2$ and feels no suppression.  An example of a sequestered state is $C^{5_1 5_2}$: although one end of the string couples to D$5_2$ the other attaches to D$5_1$ and string tension localises the string at the origin, away from the fixed point occupied by $Y_2$.
  
These two modifications are expressed in the following equations:
\begin{align}
  K(S,\overline{S},T_{i},\overline{T}_{i},Y_{2},\overline{Y}_{2})_{seq.} = & ~ \frac{1}{2} \left[ Y_{2}+\overline{Y}_{2} - \delta_{GS} \ln (T_{2} 
    + \overline{T}_{2}) \right]^{2} \nonumber \\
  & + \sum_{i \neq 2} \frac{\xi(T_{2},Y_{2})}{(S+\overline{S})} |C^{5_{i}}_{i}|^{2}  
  + \frac{1}{2} \sum_{i \neq 2} \frac{\xi(T_{2},Y_{2})}{(T_{k}
    +\overline{T}_{k})} |C^{5_{i}}_{j}|^{2}d_{ijk}  \nonumber  \\
  & + \frac{1}{2} \sum_{i} \frac{\xi(T_{2},Y_{2})}{ 
    (S+\overline{S})^{1/2} (T_{k}+\overline{T}_{k})^{1/2}} |C^{5_{i} 5_{j}}|^{2} d_{ijk} \nonumber \\
  & + \frac{1}{2} \sum_{i \neq 2} \frac{\xi(T_{2},Y_{2})}{ 
    (T_{j}+\overline{T}_{j})^{1/2} (T_{k}+\overline{T}_{k})^{1/2}} |C^{9 5_{i}}|^{2} d_{ijk} \label{eq:kahlermod}
\end{align}
and
\begin{align}
  K(S,\overline{S},T_{i},\overline{T}_i)_{unseq.} = & -\ln(S+\overline{S}) -\sum_{i=1}^3\ln(T_i + \overline{T}_i) + \sum_{i=1}^3\frac{|C_i^9|^2}{(T_i + \overline{T}_i)} \nonumber \\
  & + \frac{|C^{5_2}_2|^2}{(S + \overline{S})} + \frac{1}{2} \sum_{i}^3\frac{|C^{5_2}_k|^2}{(T_i + \overline{T}_i)}d_{ik} + \frac{|C^{95_2}|^2}{(T_1 + \overline{T}_1)^{1/2}(T_3 + \overline{T}_3)^{1/2}} \label{eq:kahler}
\end{align}
where
\begin{align}
  \xi(T_{2},Y_{2}) = exp \left[ \frac{1}{6} \left( 
      1-e^{- (T_{2} + \overline{T}_{2})\lambda_I/4} \right) 
    \left\{ Y_{2}+\overline{Y}_{2} - \delta_{GS} \ln (T_{2} + \overline{T}_{2}) 
    \right\}^{2} \right].  \label{eq:xi2}
\end{align}
The full Kahler potential is the sum of Eq.~(\ref{eq:kahlermod}) and Eq.~(\ref{eq:kahler}).

For the values of the parameters in our model, $\xi \approx 1$.  If this were not the case canonical normalisation of the Kahler potential would produce a superpotential different from Eq.~(\ref{eq:W}).

Having specified our Kahler potential we follow the procedure in
\cite{King:2001zi} and calculate the new soft terms.  They are
parametrised by Goldstino angles $\theta$ and $\phi$ and Goldstino
parameters $\Theta_i$ (where $\sum_{i=1}^3 \Theta_i^2=1$).

We now present the full expressions for the relavant soft masses, $\tilde{m}$, $A_{C^{5_1}_3 C^{5_1 5_2} C^{5_1 5_2}}$ and $A_{C^{5_2}_3 C^{5_1 5_2} C^{5_1 5_2}} $, up to $\mathcal{O}\left(\frac{1}{T_2+\overline{T}_2}\right)$

\begin{align}
  m_Q^2 & = m_{C^{5_2}_3}^2 = m_{3/2}^2 - 3m_{3/2}^2 \Theta_1^2 \cos^2\theta \sin^2\phi \nonumber\\
  m_0^2 & = m_{C^{5_{1} 5_{2}}}^{2} = \tilde{m}^{2} - \frac{3}{2} m_{3/2}^{2} 
  \left( \sin^{2} \theta + \Theta_{3}^{2} \, \cos^{2} \theta \, \sin^{2} \phi 
  \right)   \\
  m_\phi^2 & = m_{C^{5_{1}}_{3}}^{2} = \tilde{m}^{2} - \frac{3}{k} m_{3/2}^{2}  \Theta_{2}^{2} \cos^{2}\theta \sin^{2}\phi. \nonumber
\end{align}

\begin{align}
  \tilde{m}^{2} = m_{3/2}^{2} \left[ \vrule width 0pt height 18pt
    1 - \cos^{2}\theta \cos^{2}\phi
    \left( 1-e^{-\lambda_I(T_{2}+\overline{T}_{2})/4} \right) \right. \hspace*{7cm} 
  \nonumber \\
  - \frac{\cos^{2}\theta \sin^{2}\phi \, \Theta_{2}^{2} \, \delta_{GS}}{k}
  \left( 1- e^{-\lambda_I(T_{2}+\overline{T}_{2})/4} \right)
  \left\{ Y_{2}+\overline{Y}_{2}-\delta_{GS} \ln(T_{2}+\overline{T}_{2}) \right\} 
  \hspace*{1.2cm} \nonumber \\
  + \frac{\cos^{2}\theta \sin^{2}\phi \, \Theta_{2}^{2} \, 
    e^{-\lambda_I(T_{2}+\overline{T}_{2})/4}
  }{32 k} \lambda_I^2(T_{2}+\overline{T}_{2})^{2} \left\{ Y_{2}+\overline{Y}_{2}-\delta_{GS} 
    \ln(T_{2}+\overline{T}_{2}) \right\}^{2} \hspace*{0.8cm} \nonumber \\
  - \frac{\lambda_I \cos^{2}\theta \cos\phi \sin\phi \, \left( \Theta_{2} \, 
      e^{i(\alpha_{2}-\alpha_{Y_{2}})} + \Theta_{2}^{\dagger} \, 
      e^{-i(\alpha_{2}-\alpha_{Y_{2}})} \right) e^{-\lambda_I(T_{2}+\overline{T}_{2})/4} 
  }{32 \sqrt{k} }  \hspace*{0.4cm} \nonumber \\
  \times \left\{ Y_{2}+\overline{Y}_{2}-\delta_{GS} \ln(T_{2}+\overline{T}_{2}) \right\} 
  \left. \left( \vrule width 0pt height 15pt
      8 (T_{2}+\overline{T}_{2}) + \lambda_I \delta_{GS} 
      \left\{ Y_{2}+\overline{Y}_{2}-\delta_{GS} \ln(T_{2}+\overline{T}_{2}) \right\} 
      \vrule width 0pt height 16pt\right)
    \vrule width 0pt height 18pt \right].  \label{eq:mtilde}
\end{align}

\begin{align}
  A_\lambda = A_{C^{5_1}_3 C^{5_1 5_2} C^{5_1 5_2}} 
  = -\sqrt{3} m_{3/2} \, \cos\theta \left[ \sin\phi \, \Theta_1 \, 
    e^{i\alpha_1} \right. \nonumber \hspace*{1.8cm} \\
  + \sin\phi \frac{\Theta_2 \, e^{i\alpha_2}}{8\sqrt{k}} \, 
  e^{-\lambda_I(T_2 + \overline{T}_2)/4} \, \lambda_I(T_2 + \overline{T}_2)
  \left\{ Y_2 + \overline{Y}_2 - \delta_{GS} 
    \ln(T_2 + \overline{T}_2) \right\}^{2}
  \hspace*{1.5cm} \nonumber \\
  \left. - \cos\phi \, e^{i\alpha_{Y_2}}
    \, e^{-\lambda_I(T_2 + \overline{T}_2)/4} 
    \left\{ Y_2 + \overline{Y}_2-\delta_{GS} \ln(T_2 + \overline{T}_2) \right\} 
  \right] \hspace*{1cm} \label{eq:aii13}
\end{align}

\begin{align}
  A_Q = A_{C^{5_2}_3 C^{5_1 5_2} C^{5_1 5_2}} 
  = -\sqrt{3} m_{3/2} \, \cos\theta \left[ \sin\phi \, \frac{\Theta_2 \, 
    e^{i\alpha_2}}{\sqrt{k}} \right. \nonumber \hspace*{1.8cm} \\
  + \sin\phi \frac{\Theta_2 \, e^{i\alpha_2}}{12\sqrt{k}} \, 
  e^{-\lambda_I(T_2 + \overline{T}_2)/4} \, \lambda_I(T_2 + \overline{T}_2)
  \left\{ Y_2 + \overline{Y}_2 - \delta_{GS} 
    \ln(T_2 + \overline{T}_2) \right\}^{2}
  \hspace*{1.5cm} \nonumber \\
  \left. - \cos\phi \, \frac{e^{i\alpha_{Y_2}}}{3}\left(1 + 
    2e^{-\lambda_I(T_2 + \overline{T}_2)/4} \right)
    \left\{ Y_2 + \overline{Y}_2-\delta_{GS} \ln(T_2 + \overline{T}_2) \right\} 
  \right] \hspace*{1cm}
\end{align}

For our value of $\lambda_I$ it is interesting to note that, to a very good approximation, $\tilde{m}=m_{3/2}$, consistent with \cite{Ibanez:1998rf}, so the effects of the sequestering are not felt by the soft masses.  Also the exponentials vanish from $A_\lambda$ so it is not the sequestering that breaks the sum rule in Eq.~(\ref{eq:tri}).  This should not come as a surprise because $R_2$ is close to the Planck length, as we saw in Eq.~(\ref{eq:R2}), so there is next to no spatial separation between $Y_2$ and D$5_1$.  The sum rule is broken by the Kahler potential for the twisted moduli in Eq.~(\ref{eq:ykahler2}).  If it was logarithmic, as all the other moduli are, then the sum rule would hold, but the fact it is quadratic breaks the sum rule.

\end{document}